\documentclass[journal=acsnano,manuscript=article]{achemso}

\usepackage[version=3]{mhchem} %
\usepackage{xfrac}
\usepackage{amsmath}
\usepackage{enumerate}
\usepackage{amssymb}
\usepackage{gensymb}

\newcommand {\pt} [1] {\left( #1 \right)}
\newcommand {\bk} [1] {\left[ #1 \right]}
\newcommand {\cb} [1] {\left\{ #1 \right\}}
\newcommand {\abs} [1] {\left| #1 \right|}

\author{Maya M.\ Martirossyan}
\affiliation{Department of Materials Science and Engineering, Cornell University, Ithaca, NY, USA} %
\author{Matthew Spellings}
\affiliation{Vector Institute for Artificial Intelligence, Toronto, ON, Canada} %
\author{Hillary Pan}
\affiliation{Department of Materials Science and Engineering, Cornell University, Ithaca, NY, USA} %
\author{Julia Dshemuchadse}
\affiliation{Department of Materials Science and Engineering, Cornell University, Ithaca, NY, USA} %
\email{jd732@cornell.edu}

\title{Local structural features elucidate crystallization of complex structures}

\keywords{crystal growth $|$ complex structures $|$ machine learning $|$ order parameters}

\begin{document}

\begin{abstract}
Complex crystal structures are composed of multiple local environments, and how this type of order emerges spontaneously during crystal growth has yet to be fully understood. 
We study crystal growth across various structures and along different crystallization pathways, using self-assembly simulations of identical particles that interact via multi-well isotropic pair potentials. 
We apply an unsupervised machine learning method to features from bond-orientational order metrics to identify different local motifs present during a given structure's crystallization process. 
In this manner, we distinguish different crystallographic sites in highly complex structures. 
Tailoring this newly developed order parameter to structures of varying complexity and coordination number, we study the emergence of local order along a multi-step crystal growth pathway---from a low-density fluid to a high-density, supercooled amorphous liquid droplet and to a bulk crystal. 
We find a consistent under-coordination of the liquid relative to the average coordination number in the bulk crystal. 
We use our order parameter to analyze the geometrically frustrated growth of a Frank--Kasper phase and discover how structural defects compete with the formation of crystallographic sites that are higher-coordinated than the liquid environments. 
The here-presented method for classifying order on a particle-by-particle level have broad applicability to future studies of structural self-assembly and crystal growth, and they can aid in the design of building blocks and for targeting pathways of formation of novel soft-matter structures. 
\end{abstract}

\section{Introduction}
Crystal growth and nucleation constitute a rich field studied across a wide range of systems. 
Crystallization processes span length scales and materials families: from the biological---such as the crystallization of proteins in solution \cite{galkinControlProteinCrystal2000} or the assembly of gyroid structures in butterfly wings \cite{saranathanStructureFunctionSelfassembly2010}---to the physical---on the mesoscale: the organization of colloids \cite{gasserRealspaceImagingNucleation2001}, nanoparticles \cite{sunFormationPathwaysMesoporous2017}, and block copolymers \cite{songFormationPeriodicallyOrderedCalcium2016} into various structures; and on the atomic scale: the formation of quasicrystals \cite{senabulyaKineticEquilibriumShapes2018} or more ubiquitous substances such as ice \cite{matsumotoMolecularDynamicsSimulation2002}. 
The emergence of order is a ubiquitous phase transition, yet our understanding of spontaneous phenomena such as crystallization remains incomplete \cite{gasserCrystallizationThreeTwodimensional2009}. Experimental and simulation studies of growth are largely isolated to specific systems \cite{freitasUncoveringEffectsInterfaceinduced2020, smeetsCalciumCarbonateNucleation2015}, while theoretical models for studying, e.g., layer growth \cite{burtonGrowthCrystalsEquilibrium1951} or Wilson--Frenkel growth \cite{wilsonVelocitySolidificationViscosity1900, frenkelNoteRelationSpeed1932}, or methods such as kinetic Monte Carlo \cite{gilmerSimulationCrystalGrowth1972} make a number of simplifying assumptions that impede the extraction of generalized crystal growth principles for the study of more intricate crystal structures. 

The applicability of mechanisms such as particle-by-particle attachment and classical nucleation theory \cite{beckerKinetischeBehandlungKeimbildung1935, weeksDynamicsCrystalGrowth1979} has been challenged by experimental evidence across various chemistries and length scales, such as the effects of order development in liquid precursor droplets prior to nucleation \cite{vekilovDenseLiquidPrecursor2004, kaissaratosTwoStepCrystalNucleation2021b} that may enhance nucleation and growth rates \cite{sleutelRoleClustersNonclassical2014, xieSuperclustercoupledCrystalGrowth2019} or other multi-step crystallization pathways \cite{deyoreoCrystallizationParticleAttachment2015, liDirectionSpecificInteractionsControl2012, savageExperimentalEvidenceTwoStep2009, fangTwostepCrystallizationSolid2020}. 
Non-classical nucleation and crystallization theory posit that prior to nucleation, an intermediate, metastable liquid phase (or pre-nucleation cluster) is formed \cite{russoCrystalNucleationOrdering2016, walkerSolidStateTransformationAmorphous2017} lowering the barrier to nucleation \cite{tenwoldeEnhancementProteinCrystal1997}, yet a complete understanding of how liquid motifs affect crystallization remains elusive. In the case of complex crystals, growth models and experimentally observed mechanisms do not account for the presence of more than one local environment or a large periodic unit cell in the crystallizing structure. How do identical particles find their specific role to occupy in a structure with multiple local environments \cite{samantaClathratesGrow2017, doyeComputationalExplorationsSpace2021}, and what role does the liquid play in crystal growth?\cite{gebauerPrenucleationClustersSolute2014, degraafRoadmapAssemblyPolyhedral2012} As such one-component complex structures are being studied on the mesoscale \cite{linClathrateColloidalCrystals2017, reddyStableFrankKasper2018} and engineering such materials with DNA-functionalization is becoming an ever more powerful tool for design \cite{mirkinDNAbasedMethodRationally1996, macfarlaneNanoparticleSuperlatticeEngineering2011, biffiPhaseBehaviorCritical2013}, achieving a better understanding of the structural transitions and pathways during crystal growth for complex structures is critical for directing self-assembly.

Probing these questions can prove experimentally challenging, whereas simulations targeting growth of complex structures \cite{iacovellaSelfassemblySoftmatterQuasicrystals2011, keysHowQuasicrystalsGrow2007} can provide access to rich real-space positional data---the complete trajectories of all particles in the assembly. Systems of identical particles that interact via isotropic, multi-well pair potentials have been shown to spontaneously assemble into a diverse set of crystal structures \cite{engelSelfAssemblyMonatomicComplex2007, engelComputationalSelfassemblyOnecomponent2015, dshemuchadseMovingConstraintsChemistry2021,panTargetedDiscoveryLowCoordinated2023}. Such short-ranged potentials with tunable features are good models for simulating the self-assembly of soft-matter particles, resulting in structures as complex as clathrates \cite{linClathrateColloidalCrystals2017} and Frank--Kasper phases \cite{leeDiscoveryFrankKasperPhase2010, choMesophaseStructureMechanicalIonic2004, ungarFrankKasperQuasicrystalline2005}. 

\begin{figure}
\includegraphics[width=\columnwidth]{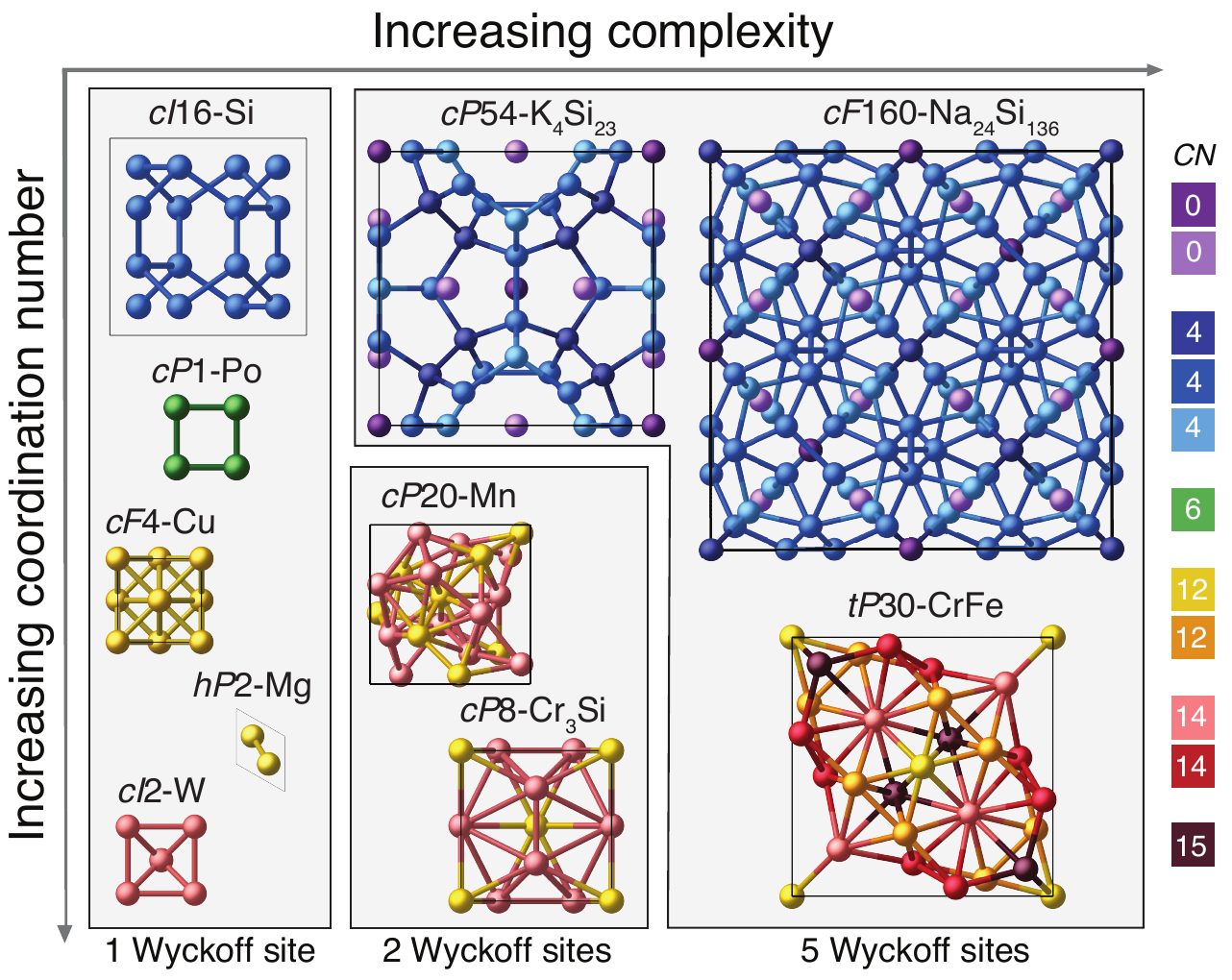}
\caption{Unit cells of structures simulated in this study, plotted by their complexity (the number of Wyckoff sites) and coordination number $CN$. Color is used to represent the coordination number ($CN$) of individual particles, ranging from purple ($CN = 0$) and blue ($CN = 4$), through green ($CN = 6$), yellow / orange ($CN = 12$), to pink / red ($CN = 14$) and maroon ($CN = 15$). Average coordination numbers for structures with more than one Wyckoff position are: $\langle CN \rangle = 3.4$ for $cP54$-K$_4$Si$_{23}$ and $cF160$-Na$_{24}$Si$_{136}$, $\langle CN \rangle = 12.8$ for $cP20$-Mn, $\langle CN \rangle = 13.5$ for $cP8$-Cr$_3$Si and $tP30$-CrFe.
}
\label{fig:coordination-complexity}
\end{figure}

In this paper, we simulate the crystal growth of model systems across a wide range of chemical interactions, ranging from low-coordinated ($CN$ = 0--6) and high-coordinated ($CN$ = 12--15) structures with varying degrees of complexity \cite{dshemuchadseMovingConstraintsChemistry2021, panTargetedDiscoveryLowCoordinated2023}, as shown in Fig.~\ref{fig:coordination-complexity}. 
We study crystallization across these structures through the evolution of local order throughout the growth process, for which we develop an order metric using unsupervised learning that can be applied robustly across various structures. 
To our knowledge, this is the first time that a large number of different local environments could be distinguished along the self-assembly pathway across a broad range of structures. We discuss the interrelation between the structure of pre-crystallization liquids and the lowest-coordinated sites in their respective crystalline solids, as well as how the liquid develops the various coordination environments of complex crystal structures in our study. Our exploration of crystal growth highlights the complexity of structural signatures present during self-assembly that depend on growth pathway as well as the presence of defect motifs that arise when a crystal has multiple local environments.

\section*{Methods}

Particle interactions adapted from previous studies were modeled with the following isotropic multi-well pair potentials: Lennard-Jones--Gauss, oscillating pair potentials \cite{dshemuchadseMovingConstraintsChemistry2021}, and a Yukawa--Gauss-based functional form \cite{panTargetedDiscoveryLowCoordinated2023}.
Few structures exist in the intermediate-coordinated ($CN$ = 7--11) range, which were therefore not included in this study. Further details of the specific interaction potential functions implemented and parameters used for assembling each structure can be found in the SI. 

Structures are labeled with their Pearson symbol, which specifies the Bravais lattice and the number of particles in the unit cell, and a prototypical compound. We denote the ``complexity'' of a structure by the number of its crystallographic Wyckoff sites, which represent symmetrically inequivalent positions in a structure's unit cell, which can therefore adopt different local environments. 

\subsection*{Simulations of crystal growth}

Molecular dynamics simulations were performed using the open-source HOOMD-blue software \cite{andersonHOOMDbluePythonPackage2020}. Systems of $N=4096$ particles were initialized with randomized velocities in an $NVT$ (canonical) ensemble at effectively zero pressure and cooled using a linear temperature ramp from $kT=1.0$ to $kT = 0.1$ controlled by a Nos\'e--Hoover thermostat. For all simulations, either $10^7$ or $10^8$ molecular dynamics time steps were used with a step size of $dt = 0.005$ to ensure that the system was being cooled sufficiently slowly for single crystals to form. The \emph{signac} and \emph{signac-flow} software were used for workflow and data management.\cite{adorfSimpleDataWorkflow2018,ramasubramaniSignacPythonFramework2018}  

Several replica simulations were performed at each state point, and simulation trajectories that exhibited low defect occurrence in the formation of high-quality crystals were chosen for generating training data for structural analysis. For each simulation, 100 total frames were collected, and training frames were chosen near (before, during, and after) the crystallization event, as well as from the final frames of the simulation. To capture more data in the crystallizing region of interest, simulations were restarted from an existing simulation frame in the temperature window where the crystallization occurred, and frames were then collected every 100 time steps. (The temperature ranges tended to be sufficiently narrow for the data volume to remain on the order of a few gigabytes per simulation trajectory for the system sizes simulated in this study.) We note that the process of restarting the simulation from an existing frame effectively creates a replica, as our simulations are not deterministic.

\subsection*{Unsupervised machine learning for local environment identification}

For quantifying local order in three-dimensional crystal structures, the Steinhardt order parameter 
\cite{steinhardtBondorientationalOrderLiquids1983},
\begin{align*}
& Q_l(i) = \bk{\dfrac{4\pi}{2l+1}\sum_{m=-l}^l \abs{q_{lm}(i)}^2}^{1/2}, \\
& \text{where } q_{lm}(i) = \dfrac{1}{\abs{N(i)}} \sum_{j\in N(i)} Y_{lm}\pt{\theta_{ij}, \phi_{ij}} \\
& \text{and } N(i) = \cb{\text{nearest neighbors of particle } i},
\end{align*}
and its neighbor-averaged variant \cite{lechnerAccurateDeterminationCrystal2008} have historically been used to fingerprint structural motifs in liquids \cite{steinhardtBondorientationalOrderLiquids1983}, distinguish liquid from crystalline order \cite{keysHowQuasicrystalsGrow2007}, differentiate between simple sphere packings (body-centered cubic (\textit{bcc}), cubic-close packed (\textit{ccp}), and hexagonally-close packed (\textit{hcp})) \cite{lechnerAccurateDeterminationCrystal2008}, distinguish structures formed in systems of hard shapes \cite{duShapedrivenSolidSolid2017}, etc. However, in all of these applications, a concerted choice of the value of $l$---the order of the spherical harmonic $Y_{lm}$---is required, which selects for the symmetries to which the parameter is sensitive. Moreover, the values of $Q_l$ are are highly dependent on the choice of neighborhood cutoff distance \cite{mickelShortcomingsBondOrientational2013}, and its per-particle values are not sufficiently sensitive to distinguish between geometrically similar Wyckoff sites in a highly complex crystal structure.

More recently, Spellings and Glotzer \cite{spellingsMachineLearningCrystal2018} created a machine learning-based method for crystal structure identification, which is able to accurately distinguish between similar complex structures using a high-dimensional dataset generated from the featurization (numerical representation) of local order in self-assembly simulations. The featurization is generated using the functional form
\begin{align*}
 & \overline{Y}_{lm}(i, k) = \dfrac{1}{k}\abs{\sum_{j\in N_k(i)}Y_{lm}(\theta_{ij},\phi_{ij})}, \\
 & \text{where } N_k(i) = \cb{k \text{ nearest neighbors of particle } i}.
\end{align*}
Values of $\overline{Y}_{lm}(i, k)$ are computed for a range of neighborhood sizes $4 \leq k \leq N_\text{max}$, and $0 \leq l \leq l_\text{max}$ values and $0\leq m \leq l$, where $N_\text{max}$ and $l_\text{max}$ are chosen according to the investigated system. Like Steinhardt's parameter, spherical harmonics are used as a set of basis functions, but a key difference lies in how rotational invariance is established. Rather than summing over values of $m$, an orientation is generated from the principal axes of inertia, which allows the separation of symmetry elements for each spherical harmonic. Such a featurization is necessarily high-dimensional and requires machine-learning methods for interpretation. Unsupervised learning approaches have shown promise in uncovering phase transitions \cite{wangDiscoveringPhaseTransitions2016, wetzelUnsupervisedLearningPhase2017} or learning atomic features \cite{reinhartUnsupervisedLearningAtomic2021, wangDescriptorfreeUnsupervisedLearning2022} from simulations.

\begin{figure*}
\includegraphics[width=\textwidth]{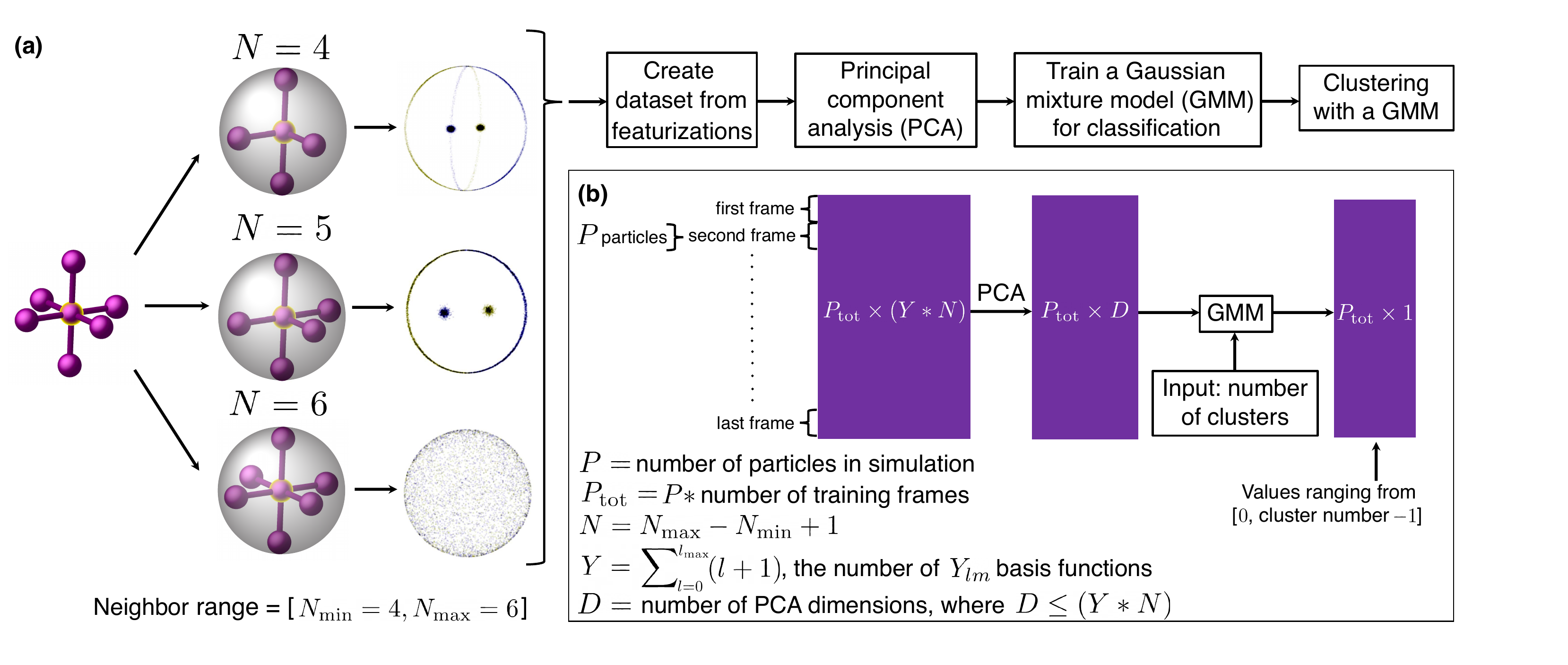}
\caption{Structural analysis pipeline for Spellings' featurization. \textbf{(a)} For each particle in the training data set, a range of neighborhood sizes is chosen (in this example, the range $[4, 6]$ is shown) and neighbors are selected based on their proximity to the central particle. The middle column illustrates the projection of the relative locations of a particle's neighbors onto a sphere. The right column shows the neighborhood's re-orientation using the principal axes of its moment-of-inertia tensor, which is done for all particles in a system snapshot, such that their superposition encompasses all possible choices of $N$ neighbors. The rotationally invariant representation shown in the right column thus also demonstrates permutation invariance---due to thermal noise, permutations of neighbor choices for the $N=4$ and $N=5$ neighborhood sizes are sampled, while the complete shell of $N=6$ particles appears featureless. \textbf{(b)} The construction of the dataset and its transformation throughout the analysis process: the featurization for each particle from each training frame is generated, and through dimensionality reduction less relevant features of the dataset are discarded. After a Gaussian Mixture Model (GMM) is trained and the number of clusters optimized, particles can be assigned to clusters, representing local environments.}
\label{fig:schematic}
\end{figure*}

We implemented both Steinhardt's $Q_l$ and Spellings' featurization via the \emph{pythia} \cite{Pythia2022} software to study the local environments present in simulations of crystallizing structures, using the features to build local order parameters. This process, as shown in Fig.~\ref{fig:schematic}, is performed on individual simulations, creating an order parameter for each crystallizing structure of interest. We applied an unsupervised learning approach in order to avoid prescribing the local environments present during crystallization a priori, as well as for ease of application to a variety of crystal structures. Similar approaches applying unsupervised methods to individual particles have been implemented with other spherical harmonics-based descriptors \cite{adorfAnalysisSelfAssemblyPathways2020, deComparingMoleculesSolids2016}, but do not use varying neighborhood sizes in their dataset construction and focus on the crystallization of simpler structures. The results presented here will describe the use of Spellings' featurization because it is higher-dimensional and contains more detailed information, but the treatment and training of the data will be nearly identical for Steinhardt's and Spellings' features (see Supplementary Information (SI)).

For all structures, $l_\text{max}=12$ was used. Typically, $N_\text{max}$ (for Steinhardt's featurization, $N = N_\text{max}$) was set to the highest coordination number expected for the idealized structure, however, in some cases it was necessary to include more neighbors. Including part of (in the case of $tP30$-CrFe and $cI16$-Si) or the entire second-neighbor coordination shell (in the case of $cP1$-Po, $cP54$-K$_4$Si$_{23}$, $cF160$-Na$_{24}$Si$_{136}$, which are all notably low-coordinated structures) showed better accuracy in distinguishing different local environments (further discussed in the Results and SI). For each structure, the smallest sufficient $N_\text{max}$ was the first free parameter fixed in the process of optimizing each structure's trained model. Additionally, we tested using subsets of neighborhood sizes rather than an entire range for Spellings' featurization, but this approach produced poor clustering results, suggesting that neighborhood sizes that did not necessarily correspond to a certain symmetry were important for fingerprinting local environments. 

Gaussian Mixture Models (GMMs) were used via the \emph{scikit-learn} software \cite{pedregosaScikitlearnMachineLearning2011} to cluster and classify particle environment data after projecting the data into 64-dimensional space using principal component analysis (PCA). For the Steinhardt featurization, since it is only 6-dimensional---one dimension per even value of $l$ ranging from 2 to 12---we did not perform PCA. The choice of the number of clusters allows for flexibility in classifying order for each structure, as the ``best'' choice would depend on the use case for which a model is trained. For example, to distinguish between non-crystalline and crystalline order, it would typically be sufficient to simply train the model using two clusters. However, for a study of the growth of complex crystals, it is desirable to train the model to distinguish between different bulk environments, the surface of the crystallite, the liquid, and the gas. Therefore, it was often necessary to use a larger number of clusters in this study. Further details for training the models are provided in the SI. We used the following criteria to select the number of clusters: (1) separation of non-crystalline environments from bulk crystalline sites, (2) separation of the liquid from the gas wherever possible, and (3) minimizing the number of clusters needed to differentiate as many local environments (Wyckoff positions) as possible, while maintaining good accuracy at low temperatures. We expect that the crystalline sites will have symmetries that may not be present in environments such as the surface, liquid, or gas, and that complex crystals with multiple local environments in the bulk should also have different symmetries at those sites.

\subsection*{Physical order metrics}

We use physically interpretable metrics to analyze the structural properties of each GMM cluster. These include coordination number, Voronoi analysis and isoperimetric quotient, local number density, and correlation functions, which are all calculated using the \emph{freud} software package \cite{ramasubramaniFreudSoftwareSuite2020}.

\subsubsection*{Coordination number}

In any discussion of ``bonding'' and nearest neighbors, it is important to consider how those neighbors are defined. For the most part, we define the first neighbor shell using a spherical cutoff at the position of the first minimum in the radial distribution function (RDF) \cite{brunnerZurAbgrenzungKoordinationssphaere1971}. The $CN$ values presented in this manuscript are calculated using a spherical cutoff at the first minimum in the RDF, unless otherwise noted. The RDF is calculated using the \emph{freud} software \cite{ramasubramaniFreudSoftwareSuite2020} and smoothed using a Gaussian filter with standard deviation $\sigma = 2$. While the RDF changes throughout the crystallization process, the locations of its maxima only change subtly so that the cutoff distance is calculated using only the last (coldest) frame of a simulation. $CN$ histograms were generated for the training frames for each GMM cluster, which served as a check for the interpretation of different clusters as representing particular Wyckoff sites. We also apply $CN$ analysis by GMM cluster to the high frame-rate crystallization data and isolate specific frames prior to and during crystallization to extract phase-specific $CN$ values.

\subsubsection*{Voronoi analysis and isoperimetric quotient}

Voronoi analysis \cite{lazarVoronoiCellAnalysis2022}---a geometric, parameter-free way of determining neighbors---is used for understanding local geometry and can serve as a proxy for bond angles, which are particularly susceptible to thermal noise for high-coordinated structures. Each neighbor in a particle's Voronoi polyhedron is weighted by the area of its corresponding facet, and these Voronoi weights (facet areas) are determined by both a neighboring particle's distance from the central particle and the locations of other nearest neighbors adjacent to the neighboring particle, which define the respective bond angles. 

For high-coordinated structures, the Voronoi polyhedra for ordered motifs correspond well to the coordination environments found using a spherical cutoff. For low-coordinated and disordered motifs, however, Voronoi polyhedra will overestimate the coordination number, by including additional faraway neighbors. We resolve this discrepancy by imposing a Voronoi weight cutoff when calculating $CN_{\text{Voronoi}}$ (here: 0.05), and for comparison, we provide figures that calculate $CN_{\text{Voronoi}}$ without cutoffs imposed in the SI. The isoperimetric quotient, $Q = 36\pi V^2 /S^3$, can also be calculated from the volume of a Voronoi polyhedron $V$ and its total surface area $S$ as a way to quantify local geometry \cite{polyaInductionAnalogyMathematics2020}. 

\subsubsection*{Local number density and correlation functions}

We define a simple correlation function $C(r) = \langle s_1(0) \cdot s_2(r) \rangle$, calculated over all particle pairs as a function of radial distance $r$, with the values of $s_1$ and $s_2$ based on the GMM cluster assignment: $C = 1$ for particle pairs both assigned to ``crystalline'' (bulk solid) clusters and $C = 0$ otherwise. This correlation function can be used to observe how crystallinity propagates during the simulation. We also use average local number density, $\langle N/V \rangle_\text{local} $, to characterize the particles in each cluster throughout the growth process.

\subsection*{Optimizing training parameters}

We study the growth process for both low- and high-coordinated structures, and in particular structures with more than one Wyckoff site, by applying our method to all 10 structures shown in Fig.~\ref{fig:coordination-complexity}. 
The optimization of training parameters, or hyperparameters, was most sensitive for low-coordinated structures, which did not contain enough structural information in their first neighbor shell. 
Highly complex structures proved challenging as well, as they were susceptible to inadvertently splitting low-density environments that were effectively identical into two clusters in order to allow for different local environments in the crystal to be distinguished. The separation of certain sites, especially with low multiplicity (i.e., number of equivalent positions in the unit cell) or highly similar neighborhood geometry or the same $CN$, as in the case of the highly complex structures, was not always possible.

The optimized hyperparameters---$N_\text{max}$ and number of clusters---are reported in Tab.~\ref{tab:envs} for each structure analyzed. The identified local environments for Spellings' featurization are labeled by inspection of simulation trajectories and by comparing physical metrics of clusters, such as coordination number. For each cluster, the ``calculated $CN$'' indicates the coordination number that was computed from the simulation trajectory, while the ``expected'' coordination number is based on the environment(s) identified as belonging to each cluster, according to the ideal crystal structure that is ultimately being assembled. 

Training time varied depending on the size of the dataset (defined by the number of training frames provided and $N_\text{max}$) as well the number of GMM clusters tested, taking a few minutes on one CPU to train a mixture model on a dataset of roughly 40,000 particles. The computational cost of cluster assignment for a system of 4096 particles was under a minute per simulation frame on a CPU, but also depended highly on the number of Gaussian clusters in the model. In the Supplementary Information (SI), we provide benchmarks for the performance of the order parameter---trained using Steinhardt's and Spellings' featurizations---against both the standard Steinhardt $Q_l$ parameter and coordination number to show the improved level of structural detail that can be captured with the here-presented method. The ability to train and apply our unsupervised models without the use of high-performance or GPU resources is a significant advantage over more sophisticated machine learning methods with fewer tuned features \cite{coliArtificialNeuralNetwork2021, wangDescriptorfreeUnsupervisedLearning2022, arobotoUniversalInterpretableClassification2023, banikCEGANNCrystalEdge2023}. Furthermore, using features based on from existing bond-orientational order metrics allows for greater interpretability of our resulting machine-learned models.

\begin{table*}
\caption{List of structures analyzed with Spellings' features: the $N_\text{max}$ hyperparameter used to train the order parameter for each structure, the number of Gaussian Mixture Model (GMM) clusters, the environments associated with each cluster, and the coordination numbers ($CN$s) of those environments (expected and calculated). The environments are described by the following: gas (low-density fluid), liquid (high-density fluid), surface(s), defects, and Wyckoff site in the bulk crystal. In almost all cases, results were highly similiar using Steinhardt's features.}
\label{tab:envs}
\centering
\resizebox{\columnwidth}{!}{\begin{tabular}{|l|r|r|l|l|l|}
\hline
\textbf{structure} & \texttt{$N_\text{max}$} & \textbf{clusters} &  \textbf{local environments} & \textbf{expected $CN$s} & \textbf{calculated $CN$s} \\ \hline

$cP54$-K$_4$Si$_{23}$ & 24 & 8 & $2 \times$ gas & - & 0 \\  
(clathrate I) & (full second shell) & & surface, liquid & - & 0--2 \\
& & & defects & - & 3 \\
& & & $24k$, $6c$ & 4, 4 & 4 \\
& & & $16i$ & 4 & 4 \\
& & & $6d$ & 0 & 0 \\
& & & $2a$ & 0 & 0  \\ %
\hline

$cF160$-Na$_{24}$Si$_{136}$ & 28 & 8 & $2 \times$ gas & - & 0 \\
(clathrate II) & (full second shell) & & surface, liquid & - & 0--2 \\ 
& & & defects & - & 3--4 \\
& & & $96g$ & 4 & 4 \\
& & & $32e, 8a$ & 4, 4 & 4 \\
& & & $16c$ & 0 & 0 \\
& & & $8b$ & 0 & 0 \\ \hline

$cI16$-Si & 11 & 4 & gas, outer surface & - & 0--1 \\
(high-pressure silicon) & (partial second shell) & & liquid, inner surface & - & 3--4 \\
  & & & $2 \times$ $16c$ & 4 & 4 \\ \hline

$cP1$-Po & 18 & 3 & gas, outer surface & - & 0--2 \\
(simple cubic) & (full second shell) & & liquid, inner surface & - & 2--6 \\
& & & $1a$ & 6 & 6 \\ \hline

$cF4$-Cu & 18 & 3 & gas, partial surface & - & 0--2 \\
(ccp/``\emph{fcc}'') & (full second shell) & & liquid, partial surface & - & 9--12 \\
& & & $4a$ & 12 & 12 \\ \hline

$hP2$-Mg & 12 & 3 & gas & - & 0--1 \\
(hcp) & (full first shell) & & liquid, surface & - & 7--12 \\
& & & $2c$ & 12 & 12 \\ \hline

$cP20$-Mn & 12 & 3 & gas, surface, liquid & - & 0--1, 6--9, 11--14 \\
($\beta$-manganese) & (partial first shell) & & $12d$ & 14 & 14 \\
& & & $8c$ & 12 & 12 \\ \hline 

$cI2$-W & 14 & 3 & gas & - & 0--2  \\
(\emph{bcc}) & (full first shell) & & liquid, surface & - & 7--14 \\
& & & $2a$ & 14 & 14 \\ \hline 

$cP8$-Cr$_3$Si & 14 & 4 & gas & - & 0--2 \\
(FK A15 phase) & (full first shell) & & liquid, surface & - & 8--14 \\
& & & $2a$, liquid & 12, - & 11--12 \\
& & & $6c$ & 14 & 14 \\ \hline

$tP30$-CrFe & 32 &8 & $2 \times$ gas & - & 0--2 \\
(FK $\sigma$-phase) & (partial second shell) & & outer surface & - & 3--11 \\
& & & inner surface, liquid, & - & 11--15 \\
& & & defects &  & \\
& & & $2b$, $8i'$ & 12, 12 & 12 \\
& & & $8i$ & 14 & 14 \\
& & & $8j$ & 14 & 14 \\
& & & $4g$ & 15 & 14--15 \\ \hline

\end{tabular}}
\end{table*}

The featurization and clustering approach outlined in this paper, with which we distinguish particles with geometrically distinct local environments, works well in application even to systems with some amount of thermal noise and disorder, such as those in our simulations. In almost all structures, the gas (i.e., low-density fluid), liquid (i.e., high-density fluid), and bulk crystal environments could be distinguished. The surface environment tended to be combined with the GMM cluster that included the liquid/high-density fluid since both motifs exhibit high degrees of anisotropy and ambiguity. The two motifs, however, can often be separated according to the time at which they occur during the simulation, because the surface environment cannot form without the presence of the crystalline bulk and is distinct from the local motifs in the dense fluid. In all crystal structures with two Wyckoff sites, the sites were separated from one another; in all crystal structures with five Wyckoff sites, all but two sites (with the same number of neighbors) were separated from one another. For the clathrate structures, we identified a GMM cluster representing 3-coordinated defects which were abundant in the training set. 

To our knowledge, no existing methods of quantifying bond-orientational order have been applied to identify many different local environments during self-assembly for a broad range of structures. The strength of this method for order classification is owed to the sensitivity of spherical harmonics-based featurizations to symmetry, because using different $l$ values and choosing neighborhood sizes prior to dimensionality reduction generates the representations that are most important for a particular crystal structure. Differentiating various crystalline sites and fluid environments, which are not necessarily known a priori, would otherwise necessitate a manual trial-and-error investigation for each crystal structure. With appropriate training parameters, the featurization captures a sufficient amount of relevant information to allow for the identification of local environments without needing supervision during training. Further details for how to train a GMM for a given structure are included in the SI. A trained model generalized well to different simulations that assembled the same crystal structure, even if those simulations were performed by modeling interactions with a different pair potential. It is worth noting that for Spellings' featurization, the variability in the data captured after PCA was typically 40--50\%, suggesting a nonlinear distribution of the original high-dimensional data that can be harnessed further in the future, using other machine learning approaches.

\section{Results and Discussion}

\begin{figure*}
\includegraphics[width=\textwidth]{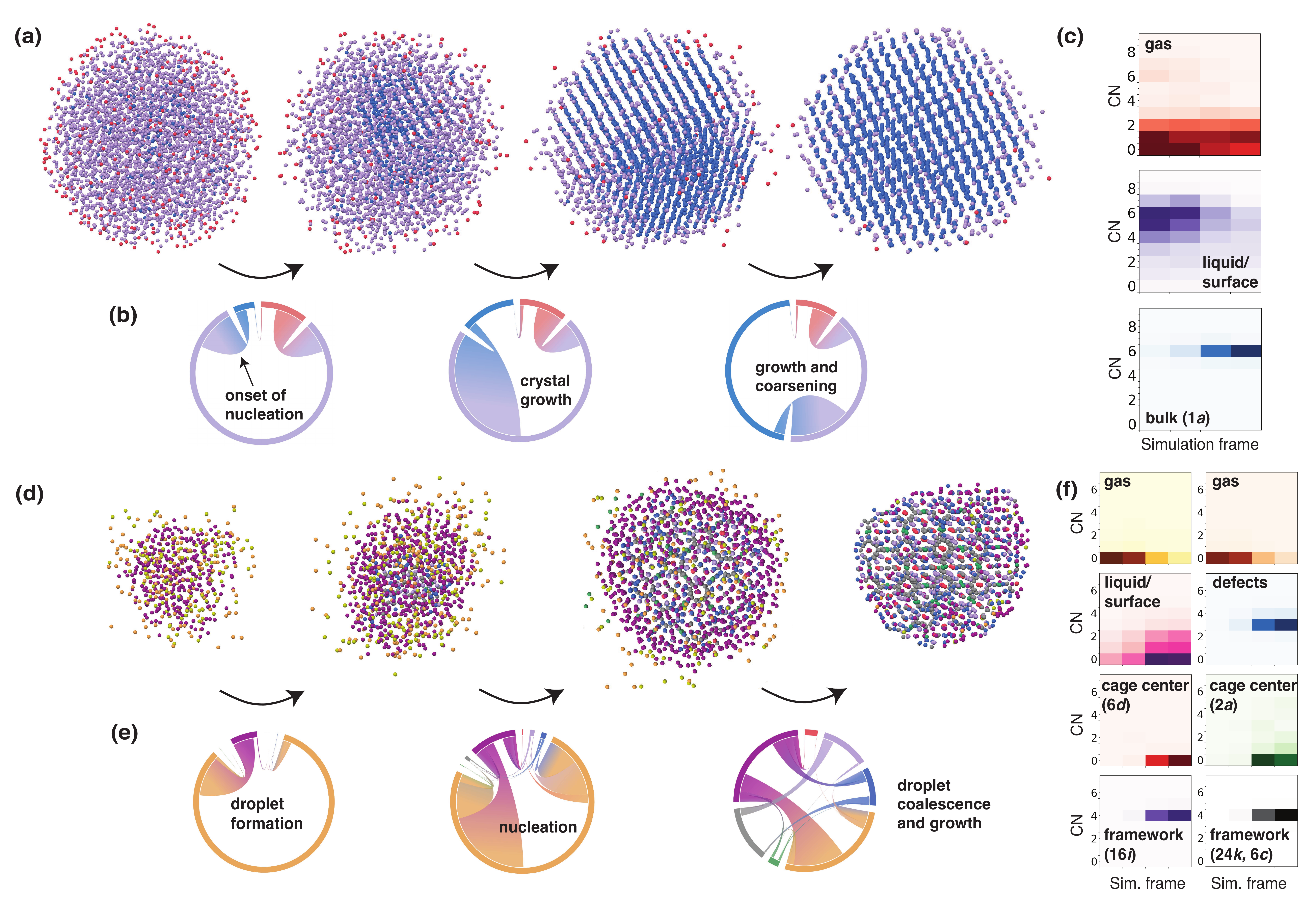}
\caption{Snapshots from simulations of growing \textbf{(a--c)} simple cubic ($cP$1-Po) and \textbf{(d--f)} clathrate-I ($cP$54-K$_4$Si$_{23}$) crystal structures colored by particles' Gaussian mixture model clusters. Visualizations in (a) and (d) are created using the \emph{flowws-analysis} \cite{spellingsFlowwsanalysis2022} software, and for ease of visualization most fluid/surface particles have been removed iteratively from the convex hull. In (a--c) the colors correspond to gas/low-density fluid and outer surface (red), liquid/high-density fluid and inner surface (lilac), and crystalline bulk (blue), while in (d--f) the colors correspond to the gas/low-density fluid (yellow and orange), three-coordinated defects (blue), surface and liquid (magenta), the 0-coordinated cage centers (green: Wyckoff site $2a$, red: Wyckoff site $6d$), and the 4-coordinated cage framework sites (lilac: Wyckoff site $16i$, gray: Wyckoff sites $24k$ and $6c$). In (b) and (e), transitions between clusters from each simulation snapshot to the next are visualized using chord plots, which show the flow between two states as well as the fraction of each cluster out of the total number of particles; in (e), the two gas/low-density fluid clusters are represented by a single grouping (orange) for simplicity. In (c) and (f), the coordination number histograms for each simulation snapshot shown in (a) and (d), respectively, are generated for each cluster and labeled accordingly.
}
\label{fig:snaps}
\end{figure*}

Our newly developed order parameter can be applied to study the growth of all ten structures shown in Fig.~\ref{fig:coordination-complexity}. For all structures, a simulation with condensation from a low-density gas phase to a dense fluid phase followed by crystallization from a dense fluid droplet was observed. The types of local environments present during this growth process are described in Tab.~\ref{tab:envs} for all crystal structures.
In Fig.~\ref{fig:snaps}, the application of the trained models to a simulation of nucleation and growth from a droplet is shown for both a simple structure ($cP1$-Po) and a complex structure ($cP54$-K$_4$Si$_{23}$), with particles colored by their GMM cluster assignment. The two examples show how differently growth can proceed from the liquid phase: in the case of $cP$54-K$_4$Si$_{23}$ (clathrate I), the dense fluid phase has barely formed when the crystalline environments become present and a crystal nucleates (apparent in the small size of the liquid droplet). We provide more detailed analysis of the growth of the clathrate I structure in the SI, and we show that the structure of the dense fluid/surface environment is structurally more similar to that of the low-density fluid. We document later in this manuscript how this is not the case for other dense fluids in our study.
In the case of $cP$1-Po (simple cubic), on the other hand, the fluid droplet constitutes the majority of particles in the simulation before nucleation and growth of the crystalline phase.

While the vast majority of our simulations crystallized via a dense fluid droplet, some exhibited growth directly from the gas phase. The latter crystallization pathway was significantly rarer across the hundreds of simulations in our dataset, likely because a dense fluid lowers the nucleation free energy barrier \cite{tenwoldeEnhancementProteinCrystal1997}. Among the three types of pair potentials used in this study, the Lennard-Jones--Gauss potential stabilized liquids with the widest temperature range, the oscillating pair potential exhibited a narrower liquid phase stability region, and the Yukawa--Gauss potential was the only pair potential to exhibit nucleation and growth directly from the gas phase---or via an exceedingly narrow liquid phase that could only be captured by high frame-rate data. 
This is likely due to the increasing steepness of Lennard-Jones--Gauss potentials vs.\ oscillating pair potentials vs.\ Yukawa--Gauss potentials. 

In several cases, we were able to compare particles interacting with different types of pair potentials that crystallized into the same structures. 
Here, we highlight the trends and observations of crystal growth across structures and growth pathways, with a particular focus on the complex structures in our study such as the Frank--Kasper phases $cP$8-Cr$_3$Si and $tP$30-CrFe. 
In the SI, we include a detailed analysis of the growth of $cI$2-W (\textit{bcc}) and $cP$54-K$_4$Si$_{23}$ (clathrate I). Application of our order parameter elucidates the structural complexity inherent in the process of crystallization and allows us to compare growth across different crystallization pathways, examine the similarity of the liquid environments to a subset of the crystalline coordination environments, and dissect the growth of a complex structure, including recrystallization.

\subsection*{Differing crystallization pathways}

For the $cP$8-Cr$_3$Si structure in particular, growth along two different pathways was observed: particle-by-particle growth directly from the gas phase was observed with the Yukawa--Gauss potential, while crystallization proceeding via a dense fluid was observed with the oscillating pair potential. This set of simulations allows us to compare the performance of the order parameter for different crystallization pathways and consider the relevance of the dense fluid to crystal growth.

We trained two models, one on simulations of each pathway. The optimal hyperparameter $N_\text{max} = 14$ reported in Tab.~\ref{tab:envs} for $cP$8-Cr$_3$Si was found by training on the simulation exhibiting growth from a high-density fluid.
Although we report an optimal cluster number hyperparameter in Tab.~\ref{tab:envs}, exploring different cluster numbers for each of the two models can elucidate the distinctiveness of various environments present during growth.
In both models, the 14-coordinated Wyckoff site ($6c$) is the first crystalline environment to be differentiated from the remaining particle environments, as early as with two clusters. 
Most of the high-density fluid and all surface particles can be separated from the 12-coordinated Wyckoff site ($2a$) in a four-cluster model, having remained combined if choosing three clusters. A second model can be trained on simulation data exhibiting growth directly from the gas phase, also with $N_\text{max} = 14$. This model behaves similarly to the first model for the three-cluster case---and differently for the four-cluster case.

\begin{figure}
\includegraphics[width=0.8\columnwidth]{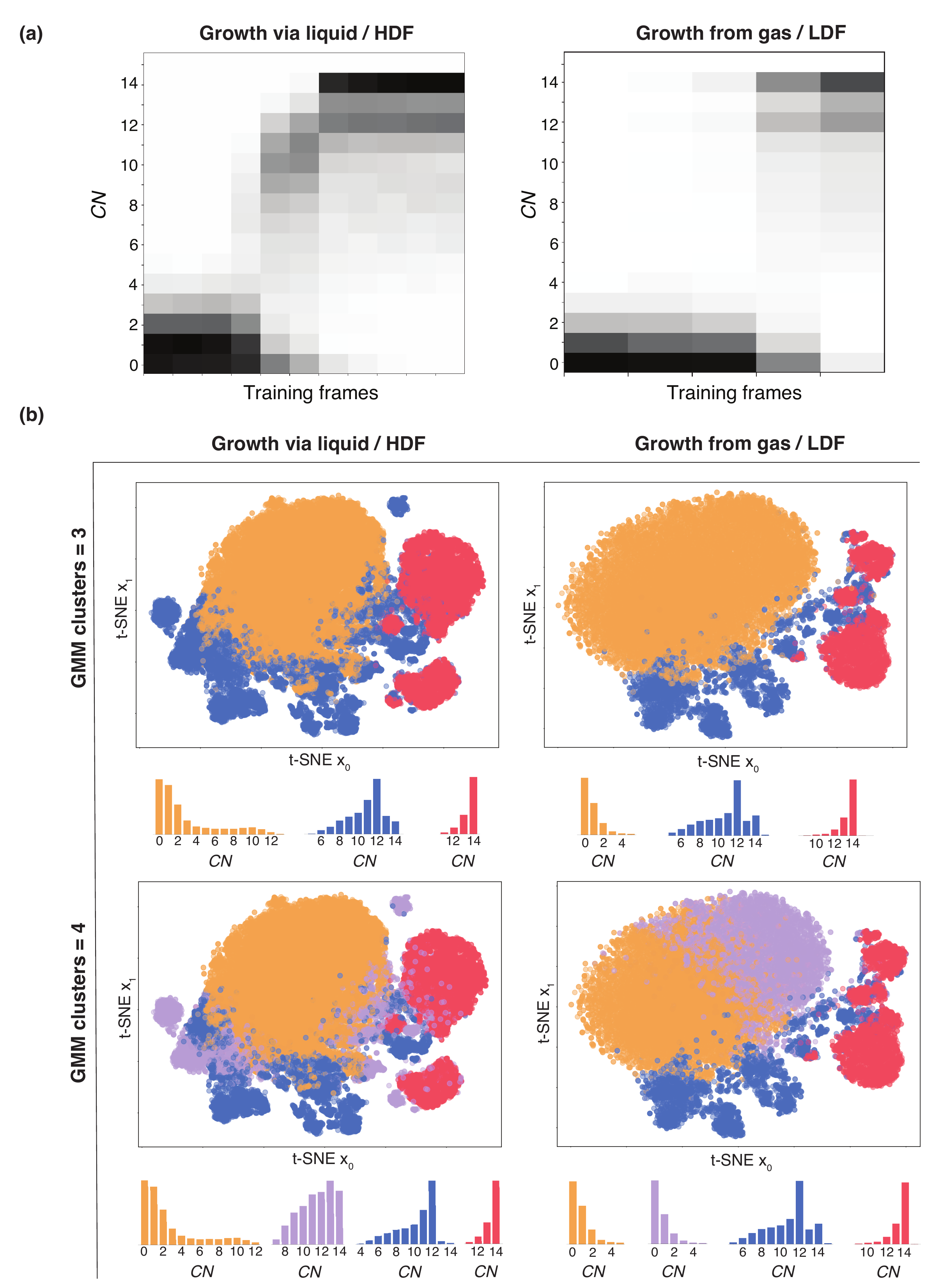}
\caption{\textbf{(a)} Coordination number histograms and \textbf{(b)} two-dimensional embeddings of particle training data from two different simulations in which $cP8$-Cr$_3$Si forms via different crystallization pathways (left: via a high-density fluid (HDF) / liquid; right: via a low-density fluid (LDF) / gas). (a) Training frames are chosen before and after crystallization in each system, and the temperature in each frame decreases from left to right. (b) Embeddings are generated using $t$-distributed stochastic neighbor embedding (t-SNE) \cite{pedregosaScikitlearnMachineLearning2011} on the PCA data and colored by each particle's GMM cluster assignment. The coordination numbers for all particle training data in each GMM cluster are shown for each model, with the histograms left unscaled for ease of visualization due to particle count differences between clusters. Detailed $CN$ histograms are provided in the SI.}
\label{fig:cP8_pathways}
\end{figure}

The application of both models is illustrated by the two-dimensional $t$-distributed stochastic neighbor embeddings (t-SNEs) of the respective training data colored by their GMM cluster in Fig.~\ref{fig:cP8_pathways}. Proximity between data points in the two-dimensional embeddings suggests structural similarity as detected by the features used. Likewise, the separation of groupings of data points in the two-dimensional embedding implies a separation in the original high-dimensional data. 
The structural dissimilarity of the 12- and 14-coordinated Wyckoff sites is apparent by the distinct separation of their GMM clusters along both growth pathways. 
In the case of growth via a liquid phase, the difference between the three-cluster and four-cluster models illustrates the structural similarity between the $cP$8-Cr$_3$Si liquid and the icosahedral environment of the 12-coordinated Wyckoff site. While icosahedral order in liquids has been demonstrated \cite{nelsonOrderFrustrationDefects1983, spaepenFivefoldSymmetryLiquids2000}, the icosahedra present in the liquid and crystalline environments are still sufficiently distinct as to be distinguished by our order parameter.

The four-cluster models can be directly compared using the $CN$ histograms for each cluster as well as the space spanned by each cluster's embedded data. In the four-cluster model trained on the simulation exhibiting growth directly from the gas phase, the fourth cluster is formed by splitting the gas environment identified in the three-cluster model in two. The two partitions exhibit equivalent $CN$ distributions, suggesting that the split is not structurally meaningful. This is unsurprising given that no dense fluid was observed in this simulation. 

For the four-cluster model trained on the simulation exhibiting growth via a dense fluid, the fourth cluster identified is structurally distinct from the other three clusters and represents most of the liquid and all surface local environments. Additional simulation snapshots and $CN$ histograms for this model are provided in the SI. 
This pair of simulations and their respective order parameters show how our simulations can model structurally distinct growth pathways, which nonetheless can lead to the formation of the same final structure. Our analysis approach and low-dimensional embeddings highlight the complexity of local structural data across a crystallization trajectory. Moreover, the embeddings shown in Fig.~\ref{fig:cP8_pathways} underscore the different relationships between the Wyckoff sites and the dense fluid phase along both growth pathways, which will further be highlighted by comparison of liquid and crystalline coordinations in the following section.

\subsection*{Liquid under-coordination}

\begin{figure}
\includegraphics[width=\columnwidth]{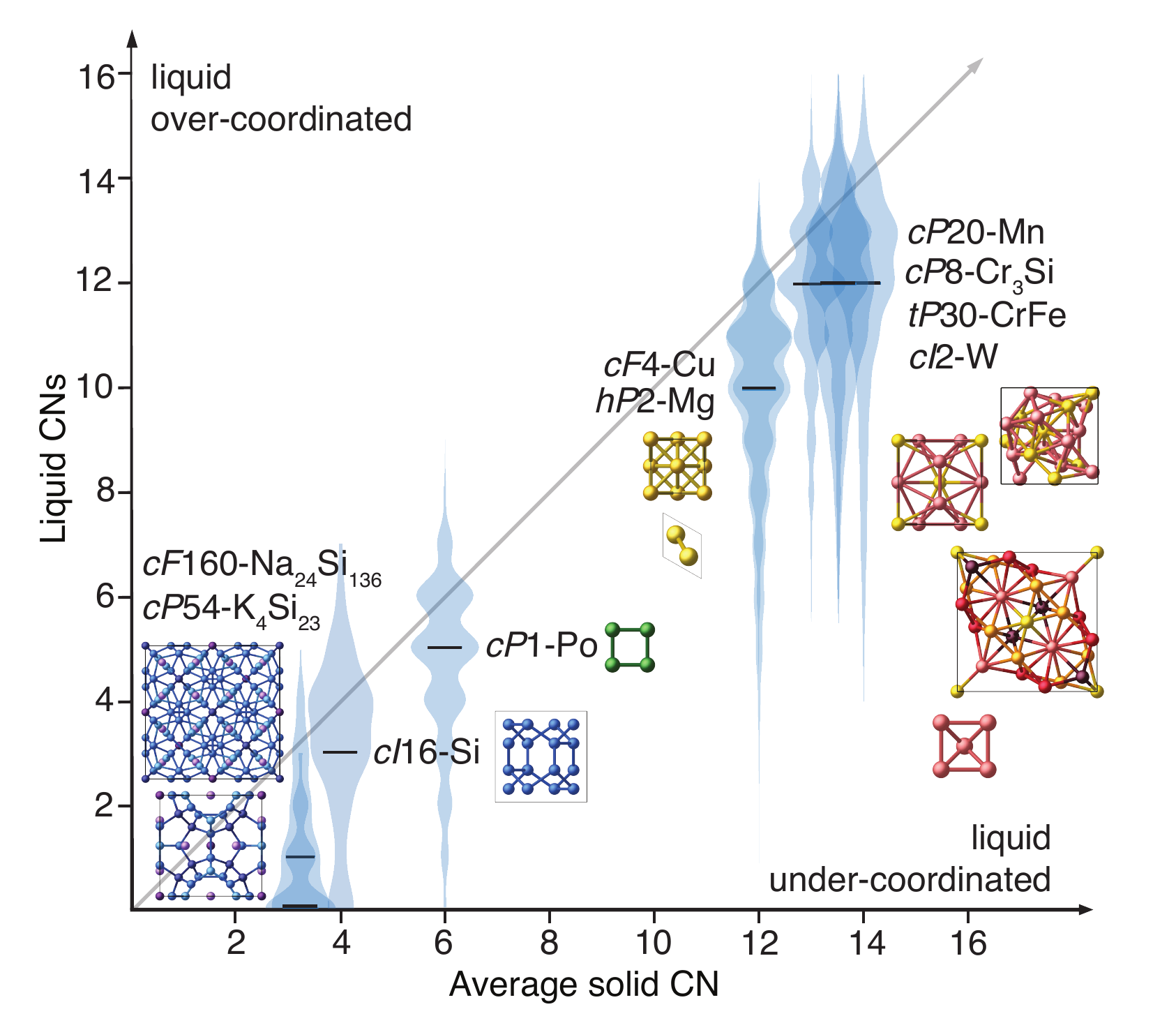}
\caption{Coordination numbers of the liquid clusters for each crystal structure plotted against that structure's theoretical average coordination number in the solid, calculated using a spherical cutoff distance from the first minumum in the solid's RDF. Violin plots are used to show the distribution of (integer) $CNs$ of the supercooled fluid immediately before crystallization, with the median marked as a horizontal line. All structures consistently exhibit under-coordination of their dense fluids relative to their bulk crystal phases.}
\label{fig:liquidCN}
\end{figure}

Short-range order in liquids of metallic elements has been studied both in simulation \cite{hafnerStructureElementsLiquid1984} as well as experiment \cite{spaepenFivefoldSymmetryLiquids2000}, identifying a relationship between the structure of the liquid and the respective solid that forms in the same system. With the multi-well potentials used in this study, we further probe how this relationship extends to the formation of complex crystal structures. We can distinguish the high-density fluid phase (i.e., liquid) from the low-density fluid (i.e., gas) and crystalline phases through classification with our order parameter(s), and we calculate $CN$s for each liquid cluster in the simulation frames immediately preceding crystallization. We find that particles in the high-density liquid-like droplet prior to crystallization has fewer nearest neighbors than the average in its solid, as seen in Fig.~\ref{fig:liquidCN}, if calculating $CN$ with a radial cutoff found from the solid's RDF. It is worth noting the diversity in coordination numbers exhibited by the pre-crystallization liquids---with respect to which our order parameter is robust---highlighting the difficulty of using straightforward $CN$ analysis to distinguish the liquid droplet from the gas (low-density fluid). 

A subtler feature, illustrated in Fig.~\ref{fig:liquidCN}, lies in the comparison of liquid under-coordination for simple structures (one Wyckoff site) to complex structures (two or more Wyckoff sites). All of the simple structures---$cI16$-Si, $cP1$-Po, $cF4$-Cu, $hP2$-Mg, and $cI2$-W---show under-coordination by one to two neighbors compared to their bulk crystals. On the other hand, the complex structures---$cF160$-Na$_{24}$Si$_{136}$, $cP54$-K$_4$Si$_{23}$, $cP20$-Mn, $cP8$-Cr$_3$Si, and $tP30$-CrFe---are not under-coordinated relative to the lowest coordination in their bulk crystal structures, but only relative to the crystalline sites with higher coordination numbers. The high-coordinated $cP20$-Mn, $cP8$-Cr$_3$Si, and $tP30$-CrFe all have a median of 12 neighbors in their dense liquid droplet, which is equivalent to the lowest-coordinated site(s) in their solids; the low-coordinated clathrates, $cP54$-K$_4$Si$_{23}$ and $cF160$-Na$_{24}$Si$_{136}$, have a median of 0 and 1 neighbor(s) in their liquids, and both have 0-coordinated cage centers in their bulk crystals.

It may seem unsurprising to report liquid under-coordination relative to each structure's bulk crystal given that a change in density is often expected upon crystallization. Prior literature on two-dimensional hard disks \cite{nelsonDislocationmediatedMeltingTwo1979} and three-dimensional Lennard-Jones liquids comprised of \textit{ccp}, \textit{hcp}, and icosahedral motifs \cite{steinhardtBondorientationalOrderLiquids1983} did not report under-coordination of the liquids for simple structures, but this discrepancy is likely due to how coordination number cutoffs are calculated (i.e., from the liquid's vs.\ solid's RDF; see SI for calculation of under-coordination for a Lennard-Jones liquid). The under-coordination of liquids has, however, been theorized for Frank--Kasper phases. These structures have multiple high-coordinated environments in their solids ($CN \geq 12$), and their liquids consist largely of icosahedral shells ($CN=12$), under- and over-coordinated particles (with $CN=10$ and $CN=14$) forming disclination lines, and pairs of particles with $CN=11$ and $CN=13$ neighbor shells signifying the presence of vacancies and interstitials, respectively \cite{nelsonOrderFrustrationDefects1983}. The formation of $CN>12$ sites upon crystallization is attributed to disclination networks forming in order to accommodate the geometric frustration from the $CN=12$ icosahedral shells present in the crystal structure. 

In more recent work, the Lennard-Jones--Gauss potential has been used to model geometrically frustrated amorphous calcium carbonate due to competing stabilizing interparticle distances from multiple potential wells\cite{nicholasGeometricallyFrustratedInteractions2023}. Geometric frustration can explain our more general observation of liquid under-coordination prior to growth of complex structures including those that are not Frank--Kasper phases, such as $cP20$-Mn ($\beta$-manganese), $cP54$-K$_4$Si$_{23}$ (clathrate I), and $cF160$-Na$_{24}$Si$_{136}$ (clathrate II). This means that the liquids of highly complex structures may be structurally more similar to only certain bulk sites, rather than all of them equally. This insight is not reflected in any existing models of crystal growth and suggests the need to incorporate the relative energetic favorabilities of possible local environments. Moreover, how the geometric frustration of local environments in the pre-crystallization liquid is resolved by crystallizing into the environments of the solid is not yet broadly understood. From our analysis of pre-crystallization liquids, it follows that the emergence of different local environments present in the complex structure must undergo different types of geometric transitions allowing them to ``choose'' their various roles in the bulk crystal.

\subsection*{Differentiation of local environments during growth and recrystallization}

Changes in local geometry for different Wyckoff sites during crystallization are investigated by applying physically interpretable metrics to our particles partitioned by their GMM cluster assignment. We investigate the growth via a dense fluid from a high frame-rate simulation of $tP30$-CrFe (Frank--Kasper $\sigma$-phase)---a structure with five Wyckoff sites in its unit cell ranging from $CN=12$ to 15. As discussed in the previous section, the Frank--Kasper liquid phases are expected to have a high occurence of $CN=12$ particles, and in this section we dissect how the higher coordinated sites ($CN=14$ and 15) emerge from the dense fluid, and how the presence of $CN=13$ defects can inhibit crystallization.

\begin{figure*}
\includegraphics[width=\textwidth]{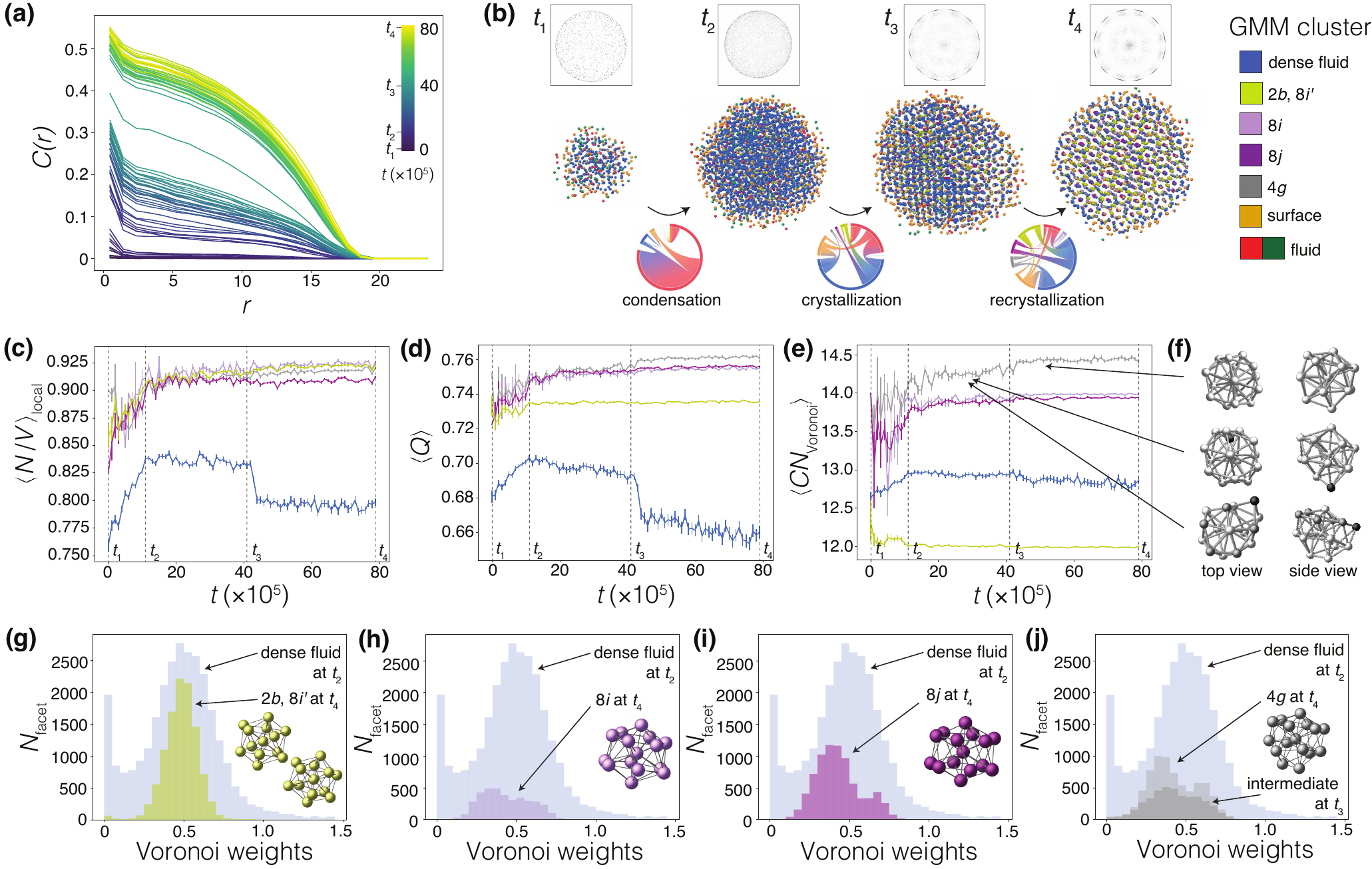}
\caption{Simulation of the growth of a $tP$30-CrFe structure type. 
\textbf{(a)} Correlation function $C(r)$ calculated as a function of interparticle distance $r$ and colored by simulation time. 
\textbf{(b)} Simulation snapshots colored by GMM cluster and bond-orientational order diagrams at times $t_1$ through $t_4$, with gas particles omitted for clarity. Switches between clusters between each time marker are represented using chord plots. 
\textbf{(c--e)} Measures of order for each cluster (excluding the gas and ``outer surface'' clusters) over time. Left to right: (c) average local number density; (d) average isoperimetric quotient of the Voronoi polyhedra; (e) average coordination number calculated from the Voronoi polyhedra using a weight cutoff. Error bars represent the standard error of the mean. 
\textbf{(f)} Representative motifs extracted from the cluster associated with the $4g$-site before and after recrystallization. The topmost row shows a typical $CN=15$ site after recrystallization, while the middle and bottom rows show $CN=14$ defects---which only appear prior to recrystallization---that are distorted $CN=15$ sites. Both of the shown defects are characterized by the presence of a square pyramidal motif formed by the inclusion of a neighbor (colored in black) outside the Voronoi weight cutoff. In the bottom row, particles in the pentagonal motif are colored dark gray as a guide for the eye. 
\textbf{(g--j)} Histograms of Voronoi weights (i.e., polyhedra facet areas) for each of the crystalline clusters from the end of the simulation (time $t_4$) plotted against the histogram of the ``dense fluid'' cluster from immediately before the onset of crystalline order (time $t_2$). The Voronoi weight histogram at time $t_3$ is included for the cluster associated with the 4g-site in (j). Idealized Voronoi polyhedra for each Wyckoff site are included as insets.}
\label{fig:voronoi_tP30}
\end{figure*}

In particular, we analyze a highly frustrated crystal growth trajectory, signified by an increased presence of $CN=13$ particles \cite{nelsonOrderFrustrationDefects1983} (see SI for $CN$ histograms). Applying GMM clustering to this simulation, coupled with Voronoi analysis, reveals that crystallization proceeds in three stages, as shown in Fig.~\ref{fig:voronoi_tP30}. Five replica simulations were run at this state point at a high frame rate, with frustrated assembly---a high occurrence of $CN=13$ during the formation of $CN=14$ and 15 sites---seen in two of the five simulations. An example of assembly where a single crystal grows directly from the fluid without such frustration is provided in the SI for comparison. 

The boundaries between the three stages illustrated in Fig.~\ref{fig:voronoi_tP30} are defined at times $t_1, ..., t_4$, as delineated by jumps in the correlation function in Fig.~\ref{fig:voronoi_tP30}a at those times. 
In the first stage, between $t_1$ and $t_2$, the droplet grows and densifies, with very few crystal-like environments appearing and disappearing via random thermal fluctuations. At time $t_2$, both isoperimetric quotient and coordination number increase for the GMM clusters associated with the $4g$ (15-coordinated), $8i$ (14-coordinated), and $8j$ (14-coordinated) Wyckoff sites. 
In the second stage, between times $t_2$ and $t_3$, densification and droplet size growth slow down and crystalline order begins to emerge, as noted by a differentiation in average local number density $\langle N/V \rangle_{\text{local}}$, average isoperimetric quotient $\langle Q \rangle$, and average coordination number $CN_{\text{Voronoi}}$, shown in Fig.~\ref{fig:voronoi_tP30}c--e. The bond-orientational order diagrams begin to show signs of global order between $t_2$ and $t_3$ (see Fig.~\ref{fig:voronoi_tP30}b), suggesting that the droplet is no longer structurally amorphous. 

At $t_3$, we see further differentiation in $\langle N/V \rangle_{\text{local}}$ for the crystalline GMM clusters (see Fig.~\ref{fig:voronoi_tP30}c), with particles in the ``dense fluid'' cluster dropping to lower $\langle N/V \rangle_{\text{local}}$ and $\langle Q \rangle$ (see Fig.~\ref{fig:voronoi_tP30}d). Also at $t_3$, the cluster associated with the 15-coordinated $4g$ site increases in $\langle Q \rangle$ (see Fig.~\ref{fig:voronoi_tP30}d) and in $CN_{\text{Voronoi}}$ (see Fig.~\ref{fig:voronoi_tP30}e). 
The final stage of assembly, from $t_3$ to $t_4$, moves the frustrated system with multiple nucleating grains and grain boundaries to a single-crystalline assembly, representing recrystallization. 
The delayed differentiation of the 15-coordinated site in the calculated physical metrics is a key difference between the frustrated and non-frustrated crystallization processes, and it is detected by the order parameter. Due to the direct crystallization of a single crystal, the non-frustrated crystallization pathway lacks the intermediate stage between time $t_2$ and $t_3$, suggesting that the proper formation of the $CN=15$ site is inhibited by the geometric frustration from $CN=13$ defects contained in the ``dense fluid'' cluster. This is not to say that the 15-coordinated site is necessarily the only one inhibited in its formation by $CN=13$ defects: the ``dense fluid'' cluster contains particles ranging from $CN=11$ to 15 as noted in Tab.~\ref{tab:envs} (and in the SI).

The particles in the GMM cluster associated with the 12-coordinated Wyckoff site(s) undergo little change in their local environment after time $t_2$, i.e., crystallization. The dense fluid cluster prior to crystallization is not under-coordinated relative to the lowest-coordinated sites in the $tP30$-CrFe crystal structure (see Fig.~\ref{fig:voronoi_tP30}e), and the distribution of Voronoi weights of the associated GMM clusters (the ``dense fluid'' cluster at $t_2$ and the $CN=12$ cluster at $t_4$) are centered at the same peak position (shown in Fig.~\ref{fig:voronoi_tP30}g--j). The Voronoi weights for the cluster associated with the $CN=12$ sites at time $t_4$ only show a narrowing of the unimodal distribution relative to that of the dense fluid at time $t_2$. By contrast, the other Wyckoff sites with $CN>12$ neighbors all exhibit geometric transitions to bimodal distributions of Voronoi weights---corresponding to the facets connected to either 5- or 6-coordinated vertices in the Frank--Kasper polyhedra \cite{frankComplexAlloyStructures1958}. In contrast, the dense fluid exhibits a unimodal Voronoi weight distribution, with the most abundant weights falling in between the two modes. We provide additional unimodal Voronoi weight data for the ``dense fluid'' cluster for the growth trajectory in the SI. The fact that the 12-coordinated sites are most similar in local structure to the dense fluid suggests that the mechanism for the emergence of 14- and 15-coordinated sites at crystallization occurs in conjunction with a reorganization of those coordination shells to accomodate more neighbors, which does not happen for the 12-coordinated crystalline sites.

At high temperature (near crystallization), our order classification method places particles into clusters associated with---but not necessarily identical to---the Wyckoff site at low temperatures, particularly because of motifs present during the frustrated assembly. This is evident in all crystalline clusters between times $t_1$ and $t_2$, for the cluster associated with the 15-coordinated Wyckoff site $4g$ between times $t_2$ and $t_3$, and vice versa for the dense fluid cluster between $t_3$ to $t_4$ (when the entire droplet has crystallized). In the case of the $CN=15$ site, we see an ``intermediate'' local environment between times $t_2$ and $t_3$, which is similar to that of the $4g$ Wyckoff site but with a lower $\langle CN_\text{Voronoi} \rangle$ value and a geometry (measured by $\langle Q \rangle$) that is more similar to that of the 14-coordinated sites. The Voronoi weight histogram of the $4g$-site associated cluster at $t_3$, as seen in Fig.~\ref{fig:voronoi_tP30}j, lacks the characteristic bimodal distribution that appears at time $t_4$. The $8i$-site associated cluster at $t_4$ is also not quite bimodal. Interestingly, the modest increase in $\langle CN_\text{Voronoi} \rangle$ for the site at time $t_3$ also vanishes if faraway neighbors (with Voronoi weights $< 0.05$) are not discarded, and instead we observe a decrease in $\langle CN_\text{Voronoi} \rangle$ at time $t_3$ for the ``dense fluid'' cluster (see SI). 

We can use this insight to extract these distorted local environments between times $t_2$ to $t_3$ from the cluster associated with the $4g$ site, and we find that---in addition to the $4g$ local environment---the cluster also detects defects, such as a distorted $4g$ sites with a pentagon and triangle motif replacing a planar hexagonal motif, as depicted in Fig.~\ref{fig:voronoi_tP30}f. As a result, the $\langle CN_\text{Voronoi} \rangle$ value of the cluster associated with the $4g$ Wyckoff site is not quite equal to 15. In the SI, we provide a further breakdown of $Q$ as it changes with $CN_{\text{Voronoi}}$ over time for the particles in this cluster and the ``dense fluid'' cluster, and we also include the Voronoi histograms for the GMM clusters associated with crystalline sites at low temperature ($kT=0.1$). The structural distortions and defects highlighted in this particular frustrated assembly and recrystallization process illuminate the need for a more structure-based approach for understanding crystal growth.

\section{Conclusion}

In this manuscript, we utilize a versatile unsupervised machine learning-based method for classifying particles by local environment in order to better understand the growth of crystal structures with different coordination numbers and degrees of complexity. Tuning the model hyperparameters in accordance with a structure's complexity and coordination number can allow for the distinction between different local geometries, which is not accomplished to the same degree by standard order metrics. This method can also distinguish between different phases---gas, liquid, and solid---even in cases where a liquid's local environments are similar to those in the solid. The broad applicability to different kinds of structures (demonstrated on 10 crystal structures with 1--160 particles per unit cells and local coordination numbers $CN=0-15$), and the ease of training a model using standard unsupervised learning methods, make our method a good local order parameter for studying the growth of complex structures in simulation. The unsupervised technique we present here can also further enable exploration of other phenomena such as surface reorganization, coarsening, and dislocation motion---in simulations or experiments. 

We use our method to analyze both fluid--fluid and fluid--solid structural transitions and extract general patterns in crystal growth. We observe under-coordination of the liquid across all structures relative to the crystal coordination, in contrast to hard anisotropic shapes, for which equal coordination number had been reported for fluids and their respective solids in simulation \cite{damascenoPredictiveSelfAssemblyPolyhedra2012}.
Our findings suggest that the geometric mechanisms for crystallization of complex structures from a dense fluid differ compared to those of simple structures, because certain crystallographic sites in a complex structure are not under-coordinated relative to their pre-crystallization liquid. 

In systems of particles interacting with isotropic pair potentials, we capture the structural transitions during crystal growth via a pathway from a low-density fluid to a high-density fluid and ultimately to a crystal, and we investigate how these transitions can differ for different Wyckoff sites in a highly complex structure. We observe the differential effect of recrystallization and frustrated assembly on specific Wyckoff sites, finding that structural defects during growth such as in $tP$30-CrFe are identified by our order parameter. Further investigation of other nucleation and crystallization pathways for complex structures is warranted, as we did not use methods such as umbrella sampling \cite{torrieNonphysicalSamplingDistributions1977} to probe difficult-to-access states near the crystallization temperature of each structure. 

The method we present could be useful for tracking nucleation in simulations with larger system sizes or as an order parameter to use in forward-flux sampling \cite{allenForwardFluxSampling2009}. Future simulation studies should also consider how the energy landscape and growth kinetics are altered by the presence of an intermediate supercooled liquid droplet. Significant work has already considered growth along different crystalline facets in simple atomic structures and even colloidal systems \cite{liMicroscopicOriginsCrystallographically2021}, or by considering liquid ordering near crystalline interfaces \cite{spaepenStructuralModelSolidliquid1975}, yet few models of crystal growth incorporate a dependence of thermodynamic quantities on local structure \cite{freitasUncoveringEffectsInterfaceinduced2020}. The structural characteristics of different local environments captured by spherical harmonics-based features could help build more sophisticated models of crystal growth. Extending growth models to explore complex structures can create a richer understanding of crystallization and aid in the design of growth units for experimental soft matter assembly in the quest to design novel materials on the mesoscale.

\begin{acknowledgement}

This material is based upon work supported by the National Science Foundation under Grant No.\ DMR-2144094 and was supported by the Cornell Center for Materials Research with funding from the NSF MRSEC program (DMR-1719875), as well as the Camille and Henry Dreyfus Foundation through a Machine Learning in the Chemical Sciences and Engineering Award (ML-22-038). M.\ M.\ M.\ acknowledges support from the National Science Foundation Graduate Research Fellowship Grant No.\ DGE-1650441 (2019--2021) and DGE-2139899 (2021--2023) and from the Dolores Zohrab Liebmann Fund Fellowship. M.\ S.\ acknowledges resources provided by the Province of Ontario, the Government of Canada through CIFAR, and companies sponsoring the Vector Institute. This work was performed at the Cornell University Center for Advanced Computing (CAC). The authors thank Rachael S.\ Skye, Reum N.\ Scott, Eric R.\ Dufresne, Alfred Amon, Benjamin Z.\ Gregory, and Samuel K.\ Bright-Thonney for helpful discussions and feedback.

\end{acknowledgement}

\begin{suppinfo}

Additional information including details of pair potentials used, discussion of how to implement the unsupervised learning method, benchmarking and comparison of the proposed method against existing metrics, details of neighbor counting, analysis of a Lennard-Jones liquid, analysis of \textit{bcc} and clathrate-I growth, and further coordination-number and Voronoi analysis examining application of the here-presented unsupervised method to Frank--Kasper phases as well as to a non-frustrated Frank--Kasper phase crystal growth trajectory.

\end{suppinfo}

\bibliography{references}

\end{document}